\documentclass[fleqn,twoside]{article}
\usepackage{graphicx}
\usepackage{espcrc2}
\usepackage[figuresright]{rotating}

\newcommand{\nn}{\nonumber}
\newcommand{\beq}{\begin{eqnarray}}
\newcommand{\eeq}{\end{eqnarray}}
\newcommand{\eq}{\ref}
\newcommand{\as}{\alpha_S}
\newcommand{\aspert}{\alpha_{S,pert}}

\newcommand{\asms}{\alpha_{{}^{\overline{\rm MS}}}}
\newcommand{\MSB}{\overline{\rm MS}}
\newcommand{\MOM}{\widetilde{\rm MOM}}

\newcommand{\lams}{\Lambda_{\overline{\rm MS}}^{Nf=2}}

\newcommand{\be}{\begin{equation}}
\newcommand{\ee}{\end{equation}}
\newcommand{\lwrsim}{\raise0.3ex\hbox{$<$\kern-0.75em\raise-1.1ex\hbox{$\sim$}}}
\newcommand{\bk}{(\beta,\kappa_{\rm sea})}
\newcommand{\ksea}{\kappa_{\rm sea}}
\newcommand{\kval}{\kappa_{\rm val}}

\def\eq#1{eq. (\ref{#1})}
\def\jhep#1#2#3{J. High Energy Phys. {\bf #1} (#2) #3}
\def\prd#1#2#3{Phys.\ Rev.\ {\bf D#1} (#2) #3}
\def\npb#1#2#3{Nucl.\ Phys.\ {\bf B#1} (#2) #3}
\def\nppb#1#2#3{Nucl.\ Phys.\ Proc. \ Suppl. \ {\bf B#1} (#2) #3}
\def\nppa#1#2#3{Nucl.\ Phys.\ Proc. \ Suppl. \ {\bf A#1} (#2) #3}
\def\plb#1#2#3{Phys.\ Lett.\ {\bf B#1} (#2) #3}

\def\prl#1#2#3{Phys.\ Rev.\ Lett.\ #1 (19#3) #2}

\def\epjc#1#2#3{Eur. \ Phys. \ J. \ {\bf C#1} (#2) #3}

\newcommand{\AmS}{{\protect\the\textfont2 A\kern-.1667em\lower.5ex\hbox{M}\kern-.125emS}}

\title{Unquenched calculation of $\as$ from Green functions: progress report.\thanks{Poster presented by H.~Moutarde\hfill\break {\tt(herve.moutarde@cpht.polytechnique.fr)}\hfill\break 
$^\dagger$ Unit\'e Mixte de Recherche  UMR 8627 du CNRS\hfill\break
$^+$Unit\'e Mixte de Recherche C7644 du CNRS}}

\author{Ph.~Boucaud\address[orsay]{Laboratoire de Physique Th\'eorique$^\dagger$, Universit\'e de Paris XI, B\^atiment 211, 91405 Orsay Cedex, France}, 
J.P.~Leroy\addressmark[orsay], 
J.~Micheli\addressmark[orsay],
H.~Moutarde\address[x]{Centre de Physique Th\'eorique de l'Ecole Polytechnique$^+$, 91128 Palaiseau cedex, France},
O.~P\`ene\addressmark[orsay], 
J.~Rodr\'{\i}guez-Quintero\address[huelva]{Dpto. de F\'{\i}sica Aplicada, E.P.S. La R\'abida, Universidad de Huelva, 21819 Palos de la fra., Spain} 
and C.~Roiesnel\addressmark[x]}

\begin{document}


\begin{abstract}
We present preliminary results on the computation of the QCD running coupling constant in the $\widetilde{MOM}$ scheme and Landau gauge with two flavours of
dynamical Wilson quarks. Gluon momenta range up to about 7~GeV ($\beta =$ 5.6,
5.8 and 6.0) with a constant dynamical quark mass. This range already allows to
exhibit some evidence for a sizeable $1/\mu^2$ correction to the asymptotic behaviour, as in the quenched approximation. We find $\Lambda_{\rm
\overline{MS}}^{N_f=2} = 264(27) {\rm MeV } \times [{a^{-1}(5.6,0.1560)}/{2.19\,
{\rm GeV}}] $, which leads to $\alpha_s(M_Z) = 0.113(3)(4)$. In view of the systematics error to be controlled, this encouraging result is more a preliminary indication than a real prediction.
\end{abstract}

\maketitle

\section{Introduction} 
From the systematic study of $\as$ in the pure Yang-Mills case \cite{green} by the Green functions method \cite{alles}, we retain that the ``asymmetric'' 3-gluon vertex (i.e. with incoming momenta ($\mu^2$,0,$\mu^2$)) is a reliable tool. We also recall that power corrections were exhibited and not negligible up to 10~GeV \cite{power}; they were eventually traced back to an $\langle A^2 \rangle$ condensate through an OPE study \cite{ope}.

This whole work is now undertaken with two flavours of dynamical quarks \cite{draft}. Following the conclusions of \cite{ope}, and denoting by $\aspert$ the perturbative running coupling up to four loops, we fit the data according to 
\be
\label{ansatz}
\alpha_{S}^{\MOM}(\mu) = \left( 1+\frac{c}{\mu^2} \right) \, \alpha_{S,{\rm pert}}^{\MOM}(\mu)
\ee 
on a large energy window obtained by combining several lattice settings corresponding to the \emph{same renormalised dynamical-quark mass expressed in physical units}. Starting from a calibrating lattice setting $\bk$=(5.6,0.1560) taken from \cite{sesam}, we search $\ksea$'s to fulfill this mass condition for $\beta$=5.8 and 6.0. To this aim, we perform an exploration of the space of bare parameters $\bk$. After that we are in position to fit the data according to $\eq{ansatz}$ and get $\lams$. At last, we derive a preliminary estimation of $\asms(M_Z)$ and briefly discuss uncertainties on this result.

\section{Exploration of parameters space $\bk$}

Given a calibrating lattice setting, we only need to evaluate lattice spacing ratios between two different sets $\bk$; this is done by ensuring the smoothness of $\as$ on the whole energy window under consideration. Quark masses are estimated using the axial Ward identity with $\ksea=\kval$
\be
\label{awi}
a m_q = \frac 1 2 \frac {Z_A} {Z_P} \frac {\sum_{\vec x} P_5(0)\partial_0 A_0(\vec x, t) }{\sum_{\vec x} P_5(0)P_5(\vec x, t)}
\ee
We have taken $Z_A$=0.77(1) and $Z_P$=0.54(1) from \cite{damir} in the RI-MOM scheme at 3~GeV, and converted $m_q$ to $\MSB$ \cite{chetyrkin}. This method provides us with first estimates (see table~\ref{runs}) which will be refined later on. We see that the masses corresponding to the lattice settings (5.6,0.1560), (5.8,0.1525) and (6.0,0.1505) are almost equal.

\begin{table}[htb]
\caption{First estimates from our runs; the masses are computed in the $\MSB$ scheme at 3 GeV.}
\label{runs}
\hspace{-0.5cm}
\begin{tabular}{@{}cccccc}\hline
$\beta$ & $\kappa_{\rm sea}$ & Volume & $a^{-1}$ (GeV) &
 $m_{\rm sea}$ (MeV)
\\ \hline
 5.6 & 0.1560 & 24$^4$/16$^4$ & 2.19(8) &  164(7)\\
 5.6 & 0.1575 & 16$^4$ & 2.42(9) & 79(3) \\
 \hline
 5.8 & 0.1500 & 16$^4$ & 2.45(13) & 325(18)\\
 5.8 & 0.1525 & 16$^4$ & 2.76(7) & 173(4)\\
 5.8 & 0.1535 & 16$^4$ & 2.91(18) & 103(16)\\
 5.8 & 0.1540 & 16$^4$ & 3.13(13) & 64(5)\\
 \hline
 6.0 & 0.1480 & 16$^4$ & 3.62(10) & 391(12)\\
 6.0 & 0.1490 & 16$^4$ & 3.73(13) & 308(12)\\
 6.0 & 0.1500 & 16$^4$ & 3.78(14) & 213(3)\\
 6.0 & 0.1505 & 16$^4$ & 3.84(15) & 169(8)\\
 6.0 & 0.1510 & 16$^4$ & 3.96(16) & 96(4)\\
 \hline
\end{tabular}\\[2pt]
\label{table_calc}
\end{table}

Because of the computational cost of this explorative stage, we only worked on small volumes. However two checks indicate that the information we gathered is relevant to tune $\bk$ for runs on larger volumes. Firstly we did not observe any significant differences on $\as$ at large momenta between our two runs at $\bk$=(5.6,0.1560). Secondly there could be a deconfinement and~/~or chiral phase transition at very small volumes, invalidating the relation $m_{P}^2 \propto (m_{q}+m_{\bar{q}})$, where $m_P$ denotes the mass of the lightest pseudoscalar fitted on our lattices, or, strictly speaking, the pion mass mixed (due to the small lattice length) with some excited states. We found empirically that 
\be
m_P^2 = 2B m_{\rm sea} + \frac{r}{V}
\ee
with $B=2.74(5)$~GeV and $r=1.41(5)$~GeV$^2$fm$^4$ from a best fit with a $\chi^2/{\rm d.o.f.}$=0.57. This relation indicates that we suffer from strong but smooth finite volume effects on $m_P$ (which is an IR-sensitive quantity). We have nevertheless some control on it, since at the continuum limit at $\bk$=(5.6,0.1560) we recover $m_{P,\infty}^2 \simeq 0.432(18)~{\rm GeV}^2$, in good agreement with SESAM \cite{sesam}; furthermore, we get $(m_u+m_d)/2 \simeq 3.6$~MeV from the pion mass and $m_s \simeq 90$~MeV from the kaon mass. This compares fairly well to other lattice estimates.

\section{Fitting $\lams$ and power corrections}

\begin{figure}[hbt]
\begin{center}
\includegraphics[height=5cm]{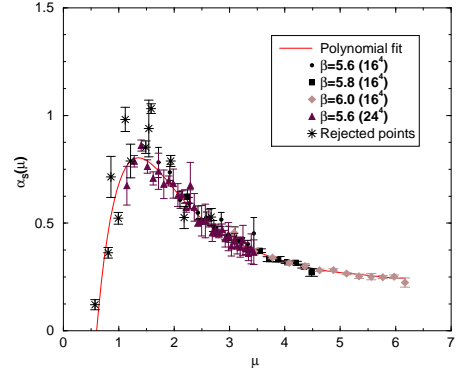}
\vspace{-1.2cm}
\caption{Matching of lattice spacings.}
\label{polynome}
\end{center}
\end{figure}

Keeping the set of lattice settings with constant dynamical mass, we refine our estimations of lattice spacings by considering a polynomial fitting all the lattice data (see figure~\ref{polynome}) except few points corresponding to $n=(L\mu)^2/(4 \pi^2) \lwrsim 2-4$ (where $L$ is the length of the lattice), presumably affected by strong  finite volume effects \cite{green}. The stability of the fit is fairly good and yields to
\be
\begin{array}{lcl} 
a^{-1}(5.8, 0.1525) & = & 2.85 \pm .09 \pm .04 \nn \\
& & \times  \frac {a^{-1}(5.6,0.1560)}{2.19\,{\rm GeV}} \ {\rm GeV} \\
a^{-1}(6.0, 0.1505) & = & 3.92 \pm .11 \pm .07 \nn \\ 
& & \times  \frac{a^{-1}(5.6,0.1560)}{2.19\,{\rm GeV}} \ {\rm GeV}
\end{array}
\ee
The central value  corresponds to a cut at $n \ge 3$,  the first error is statistical and the second is  systematic. We are going to use these sharper values from now on. Since the error on the calibrating lattice spacings propagates trivially, we drop it till the discussion of systematics.

\vspace{-0.5cm}
\vspace{-0.5cm}

\begin{figure}[hbt]
\begin{center}
\includegraphics[height=5cm]{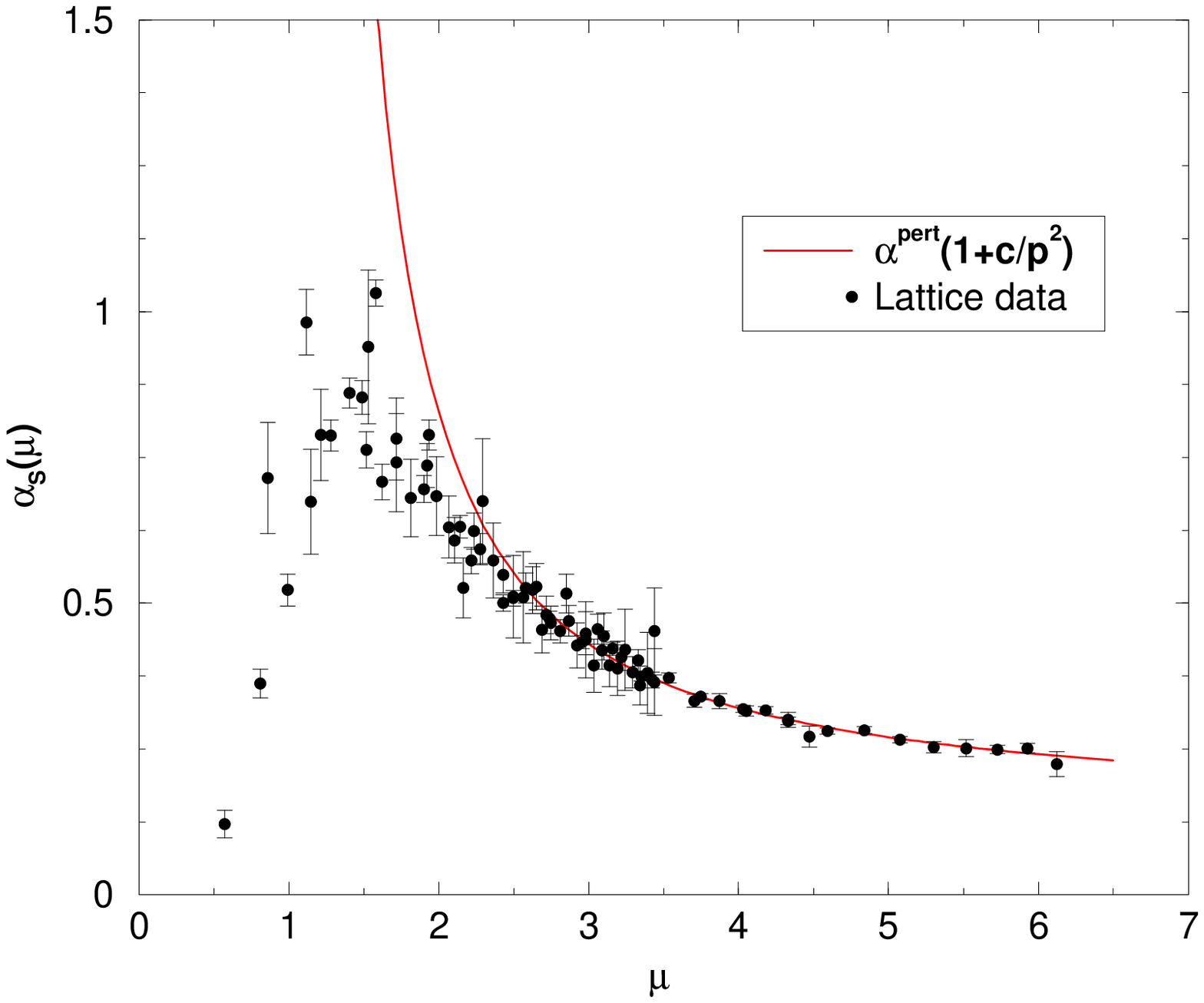}
\vspace{-1.2cm}
\caption{Asymptotic fit of $\as$.} 
\label{asymptotic}
\end{center}
\end{figure}

Having fixed the lattice spacings, we fit $\lams$ according to $\eq{ansatz}$ on a wide energy window ranging from 2.6~GeV, giving (see figure~\ref{asymptotic})
\beq 
\Lambda_{\rm \overline{MS}}^{N_f=2} & = & 264(27)\frac
{a^{-1}(5.6,0.1560)}{2.19 {\rm GeV}} {\rm MeV } \\
c & = & 2.7(1.2) \left[\frac {a^{-1}(5.6,0.1560)}{2.19\ {\rm GeV}} {\rm GeV}\right]^2 
\eeq
with a reasonable $\chi^2/{\rm d.o.f.} \simeq 0.6$.

At this point, we should stress that another fit \emph{without} power corrections is also acceptable ($\chi^2/d.o.f. = 0.96$), giving 
$\Lambda_{\rm \overline{MS}}^{N_f=2} = 345(6)$~MeV. Therefore more statistics is needed to determine the size of power corrections. 

\section{Running of $\as$ to $M_Z$ and discussion}

Concerning finite spacing effects, let us first remark that the use of non-improved dynamical quarks leads to $O(a)$ errors. This induces only a slight effect on the ratios of our \emph{small} lattice spacings. The dominant error thus comes from the calibrating lattice spacing and propagates multiplicatively. From \cite{edwards} and \cite{sesam}, we estimate this systematic uncertainty to be $\simeq$ 50~MeV on $\Lambda_{\rm \overline{MS}}^{N_f=2}$; its effect on $\alpha_s(M_Z)$ is an error of $\pm .004$. The smooth junction of $\as$ from different lattices show that other overwhelming effects are absent. 

We neglect of course power corrections in the extrapolation to $M_Z$.
If we perform the same analysis (i.e. according to $\eq{ansatz}$) with quenched data, we can extrapolate our $N_f=0,2$ results to $N_f=3$ at, say,  the charm threshold. A standard running procedure \cite{pdg}, taking the charm and beauty thresholds at 1.3~GeV and 4.3~GeV, gives 
\beq
\as(M_Z) & = & 0.113(3)(4) \\
\aspert(M_{\tau}) & = & 0.283(18)(37)
\eeq 
The second error comes from the error on the calibrating lattice spacing. This result is 2$\sigma$ below the world average experimental result $\as(M_Z) = 0.119(2)$ and the result of ALEPH \cite{aleph} $\as(M_\tau) = 0.334(22)$. It is compatible with the recent result of the QCDSF-UKQCD coll. \cite{schierholz}; older results using NRQCD were closer to experiment \cite{davies}.

Being 2$\sigma$ below the experimental value is very encouraging. Work is now in progres to check in more details the effects of our large quark masses and small lattices. This will allow us to refine our preliminary result on $\as(M_Z)$ in a near future.

\section*{Acknowledgements}

This work was supported in part by the EU 5th FP under contract HPRN-CT-2000-00145.


\end{document}